\documentclass[final,3p,times,12pt]{elsarticle}

\usepackage{graphicx,graphics}
\usepackage{amssymb,epsfig,amsmath,comment,hyperref,verbatim,color}
\usepackage{tipa}
\usepackage{multirow,url,tabularx}
\usepackage{amsfonts,comment}
\usepackage{array,bm,enumerate}
\usepackage{algpseudocode}
\begin{document}

\begin{frontmatter}
\journal{Computer Speech and Language}

\title{Refining a Deep Learning-based Formant Tracker using Linear Prediction Methods}

\author{Paavo Alku$^1$, Sudarsana Reddy Kadiri$^1$, Dhananjaya Gowda$^2$}
\address{$^1$Department of Information and Communications Engineering, Aalto University, Finland\\$^2$Samsung Research, South Korea}

\cortext[mycorrespondingauthor]{Corresponding author (Sudarsana Reddy Kadiri; sudarsana.kadiri@aalto.fi)}

\begin{abstract}
In this study, formant tracking is investigated by refining the formants tracked by an existing data-driven tracker, DeepFormants, using the formants estimated in a model-driven manner by linear prediction (LP) -based methods. As LP-based formant estimation methods, conventional covariance analysis (LP-COV) and the recently proposed quasi-closed phase forward-backward (QCP-FB) analysis are used. In the proposed refinement approach, the contours of the three lowest formants are first predicted by the data-driven DeepFormants tracker, and the predicted formants are replaced frame-wise with local spectral peaks shown by the model-driven LP-based methods. The refinement procedure can be plugged into the DeepFormants tracker with no need for any new data learning. Two refined DeepFormants trackers were compared with the original DeepFormants and with five known traditional trackers using the popular vocal tract resonance (VTR) corpus. The results indicated that the data-driven DeepFormants trackers outperformed the conventional trackers and that the best performance was obtained by refining the formants predicted by DeepFormants using QCP-FB analysis. { In addition, by tracking formants using VTR speech that was corrupted by additive noise, the study showed that the refined DeepFormants trackers were more resilient to noise than the reference trackers.} In general, these results suggest that LP-based model-driven approaches, which have traditionally been used in formant estimation, can be combined with a modern data-driven tracker easily with no further training to improve the tracker's performance.    
\end{abstract}
\end{frontmatter}

\section{Introduction}
\label{sec:intro} 
Resonance frequencies of the vocal tract, formants, are among the most important parameters of speech signals. In continuous speech, formants vary over time, manifesting themselves as time-domain contours. Formant contours have been investigated in many studies in different areas of speech science, such as in acoustic phonetics \cite{Assmann1995, Hillenbrand1995}, hearing research \cite{Schilling1998,Bruce2004} and in analysis of pathological speech \cite{Rusz2013, Kent1999}. In order to automatically estimate formant contours from speech signals, formant tracking technology is needed. Formant tracking is a challenging engineering problem, and therefore many methods have been proposed over the past few decades to track formants \cite{praat2001,wavesurfer2000,lideng2007,Mehta2012,Story2016}.
These techniques typically consist of two parts. In the first part (the estimation stage), initial values of the formants are computed in short timeframes (e.g., 25 ms) using linear prediction (LP) \cite{makhoul1975} or cepstral analysis \cite{Oppenheim2004}. 
In the second part (the tracking stage), the formants extracted in individual frames in the estimation stage are expressed using contours that cover a longer speech unit (e.g., a syllable, word or sentence) \cite{praat2001,wavesurfer2000}.
Moreover, in some techniques the estimation and tracking stages are computed simultaneously using an initial representation of the vocal tract system \cite{lideng2007,Mehta2012}.

Formant trackers can be divided into two categories based on the technology that is used in the formant estimation stage. The first category consists of classical trackers whose formant estimation is based on $model$-$driven$ signal processing methodology. In this tracker category, all-pole spectral estimation methods based on different variants of LP are mostly used to estimate formants from short timeframes of speech. Formant estimates are generally obtained in these methods either by peak-picking the power spectrum of the parametric all-pole spectral model computed by the underlying LP-based method \cite{Hillenbrand1995, Hagiwara1997} or by solving the roots of the denominator polynomial of the all-pole model \cite{Rahman2007, Wang2019}. In model-driven LP-based trackers, importantly, estimation of formants is computed directly from the test speech utterance using  the underlying LP-based signal processing algorithm without training the model using formant data. As an alternative to the model-driven formant estimation approach, a few recent formant tracking studies have used the $data$-$driven$ formant estimation approach. This approach corresponds to first training a deep learning (DL) neural network model to map selected acoustic features to formants, and then estimating formants from test utterances by computing the selected features from speech and by feeding them as input to the network. In the next two paragraphs, a brief literature review is given on some of the previous investigations representing the model-driven and data-driven approaches in formant estimation.

The most popular classic model-based approaches used in formant estimation are the autocorrelation and covariance methods of LP 
\cite{praat2001,wavesurfer2000}. Closed phase (CP) analysis is known to improve formant estimation accuracy by avoiding the contribution of speech samples in the open phase of the glottal cycle and thereby decoupling the effect of the trachea more effectively \cite{Yegna1998}. CP analysis, however, works better for low-pitched male voices, which typically have a larger number of samples in the closed phase of the glottal cycle compared to high-pitched voices of women or children, which might have just a few samples in the closed phase. To reduce problems caused by having a small number of closed phase samples, LP can be computed over multiple neighboring cycles \cite{Yegna1998}. The LP-based estimation of formants has also been studied based on all-pole phase spectra or on combinations of all-pole phase and amplitude spectra \cite{Yegna1978, Murthy2011, Vijayan2019}. Weighted linear prediction (WLP) is another example of an LP-based method that has been used in formant estimation   \cite{Lee1988,ma1993,PaavoJASA2013,Manu2014}. 
WLP is based on temporally weighting the prediction error in LP, an approach that has been shown to be beneficial in computing vocal tract models that are robust with respect to noise \cite{Lee1988,ma1993} and the biasing effect of high fundamental frequency \cite{PaavoJASA2013}. In \cite{QCPFB_JASA}, a WLP-based method called quasi-closed phase forward-backward analysis (QCP-FB) was proposed and the method was shown to outperform five reference methods in formant estimation. The improved formant estimation accuracy of QCP-FB is due to the following properties of the algorithm: (1) by using temporal weighting of the prediction error (the residual), QCP-FB is able to downgrade the effect of the glottal source in the estimation of formants, and (2) by using forward-backward (FB) analysis, the number of speech samples can be increased in LP by using two prediction directions simultaneously. 
 
An example of the data-driven approach is the formant tracking study in \cite{Dissen2019}, which investigated two DL models in formant estimation (multi-layer perceptron (MLP) and convolutional neural network (CNN)). The DL models were trained using supervised learning based on the manually annotated vocal tract resonance (VTR) speech corpus \cite{lideng2006}. A similar formant estimation method based on supervised learning was studied in \cite{Dai2020} using a bilinear network and a temporal attention-augmented bilinear network as DL models to predict formants. The same authors continued their formant tracking studies in \cite{Dai2020IS} using dilated CNNs that were trained in a supervised manner with the VTR corpus. In addition, an unsupervised DL-based formant tracker that requires no prior formant measurements as training data was studied recently in \cite{Lilley2021}. Their method uses an autoencoder type of DL network whose latent features are interpreted as formants via a special loss function. In \cite{Shrem2022}, formant tracking was studied using a CNN that maps a spectrogram into a latent representation without supervised training. The latent representation was then processed by multiple decoders, each responsible for predicting a different formant while considering the lower formant predictions. 

In the current study, formant tracking is studied by combining the model-driven and data-driven approaches. The combination, called the $refinement$ of a data-driven formant tracker, is based on first tracking formants from an utterance using an existing DL-based tracker. The tracked formants are then refined by replacing them with the formants predicted frame-wise by a model-driven, LP-based signal processing approach. By combining the model-driven and data-driven approaches we aim to tackle the following two known problems of formant estimation. (1) Data-driven methods suffer from over-fitting the formant estimation model to the training data and therefore estimation accuracy deteriorates for unseen test data \cite{Shrem2022}. We hypothesize that the decrease in formant estimation accuracy caused by over-fitting to the training data could be reduced by refining the predicted formants using a model (such as LP or QCP-FB) that is free from data learning. (2) Model-driven formant estimation methods based on different modifications of LP suffer from spurious peaks in all-pole spectra that occur, for example, in the covariance method when the all-pole model order is large relative to the number of speech samples in the covariance frame \cite{Kay1981}. The problem caused by spurious peaks can be avoided in the proposed method by taking into account only those peaks that occur closest to the formants detected by the DL-based tracker (as will be described in more detail in Section \ref{sec:methods}).

The combined use of model-driven and data-driven approaches was studied for the first time by the present authors in \cite{Gowda2021}. The current study extends our previous investigation in two ways. First, the preliminary study published in \cite{Gowda2021} reported results of formant tracking experiments only for vowels, diphthongs, and semivowels, whereas the experiments of the current study are reported also for more fine-grained phonetic categories. Second, and more importantly, our preliminary study published in \cite{Gowda2021} used a simple deep neural network (DNN) -based model to map speech features into formant estimates. {The DNN was trained for the purposes of the preliminary study only, and the model has not been used in any other formant tracking experiments except in those that are reported in \cite{Gowda2021}. In the current study, however, we study a more general scenario in which formant contours predicted by an $existing$ default DL-based formant tracker are refined using formant information obtained from the model-driven approach. We consider this general refinement scenario important because its investigation addresses the following interesting question that has not been studied in the area of formant tracking before: Can formant tracking accuracy of an existing, modern data-driven tracker be improved by plugging into the system an efficient model-driven signal processing-based module, and by refining the formants predicted by the data-driven tracker using formant information estimated by the model-driven module? As the existing data-driven tracker, we selected the DeepFormants tracker \cite{Dissen2019} which, to the best of our knowledge, is the first formant tracker where DL-based formant estimation and tracking is used and which therefore can be regarded as the default system in the category of data-driven formant trackers. In addition, DeepFormants is the only modern data-driven tracker that is publicly available (\url{https://github.com/MLSpeech/DeepFormants}), which makes it an ideal tracker to study the proposed refinement approach.} 

The goal of the current study is to investigate the accuracy of formant tracking based on refining the formant contours predicted by DeepFormants using two LP-based model-driven methods, conventional LP and QCP-FB. The DeepFormants trackers based on the refinement are compared with the original version of Deepformants \cite{Dissen2019}, which uses no refinement, and with five known traditional baseline trackers that all use model-driven LP-based methods in formant estimation. The main novelty is in studying whether the accuracy of the DeepFormants tracker, which has been trained as reported in \cite{Dissen2019}, can be improved with $no$ further training by refining the predicted formant contours using formant information extracted from model-driven signal processing methods. 

The proposed method to refine the formants estimated by DeepFormants is described in Section \ref{sec:methods} by first giving a brief overview of the DeepFormant tracker and two LP-based methods that are used in the refinement of the tracker. In Section \ref{sec:experiments}, the experimental setup of our formant tracking experiments is described. The results of the study are reported in Section \ref{sec:results}. {The main results are discussed, and conclusions are drawn in Section \ref{sec:conclusions}.}

\section{Methods}
\label{sec:methods} 
In this section, the methodological background of the proposed refinement approach is described in Section \ref{sec:background} by first summarizing the main properties of DeepFormants and then describing two LP-based methods that were selected to be used as model-driven formant estimation methods in refining formant contours predicted by DeepFormants. After this, the proposed refinement technique in formant tracking is described in Section \ref{sec:refining}.
\subsection{Background methods}
\label{sec:background} 
\subsubsection{The DeepFormants tracker}
\label{sec:deepformants} 
As the default data-driven formant tracker, the current study uses DeepFormants whose early version was published in \cite{Dissen2016} and the tracker was later extended in \cite{Dissen2019}. The DeepFormants tracker used in the current study is based on \cite{Dissen2019} and we used the implementation available at \url{https://github.com/MLSpeech/DeepFormants}. DeepFormants is a data-driven formant tracker consisting of an estimation stage and a tracking stage, both of which are implemented using neural networks (NNs) that are trained with supervised learning based on manually annotated formant data. In \cite{Dissen2019}, two feedforward NNs (MLP and CNN) were used in the estimation stage. For the tracking stage, \cite{Dissen2019} studied two recurrent NN architectures, the long short-term memory (LSTM) network and the convolutional recurrent network. In the current study, we use the DeepFormants version that uses the MLP model and the LSTM model in the estimation and tracking stage, respectively.

In the estimation stage, the input to DeepFormants is a vector of 350 features computed from speech. The feature vector consists of cepstral features computed from LP filters of different orders and features of the quasi-pitch-synchronous speech spectrum. The output of the network is a vector corresponding to the first ($F_1$), second ($F_2$), and third ($F_3$) formant to be predicted. The network has three fully connected hidden layers with 1024, 512, and 256 neurons, and the sigmoid function is used for activation. The network was trained in \cite{Dissen2019} based on the regression task using Adagrad \cite{Duchi2011} and the mean absolute error criterion between the predicted formant and its ground truth. All three formants are predicted simultaneously by the network. 
For the tracking stage, the DeepFormant tracker used in the current study utilizes an RNN consisting of an input layer with the same 350 input features as in the estimation stage. After the input layer, the network has two LSTM layers with 512 and 256 neurons, a time-distributed fully connected layer with 256 neurons, and an output layer consisting of the three formant frequencies. Similarly to the NN used in the estimation stage, the sigmoid function is used in activations and the model is optimized using Adagrad based on the mean absolute error. {The DeepFormants tracker was trained in \cite{Dissen2019} using the training set of the VTR database described in \cite{lideng2006}.  The VTR database contains altogether 516 utterances selected from the popular TIMIT database \cite{timit}. The training set of VTR contains 324 utterances that were produced by 162 speakers (97 males, 65 females) each producing two sentences (one phonetically compact sentence and one phonetically diverse sentence).}

\subsubsection{The selected LP-based formant estimation methods}
\label{sec:lpmethods} 
Two LP-based methods were used as model-driven formant estimation approaches to refine the formants tracked by DeepFormants. The first, LP based on the covariance method (LP-COV), was chosen to represent classic LP-based all-pole spectral modeling techniques \cite{makhoul1975} that have been widely used in formant estimation studies. The second one, QCP-FB, was selected because it showed the best performance in a comparison of six model-driven formant estimation methods in \cite{QCPFB_JASA}. LP-COV is an established method in speech processing and therefore the reader is referred to \cite{makhoul1975} for a detailed description of it. A brief description of QCP-FB, however, is given below.

The traditional formulation of LP, which is used, for instance, in LP-COV, is based on forward prediction in which the current speech sample is predicted from the past $p$ samples. It is, however, also possible to use backward prediction in which the current sample is predicted from the future $p$ samples.
The combination of these two, forward-backward (FB) analysis, is used in QCP-FB. 
The combined error to be minimized is given by
\vspace{-2mm}
\begin{equation}
\vspace{-2mm}
{\cal E}={\cal E}^f+{\cal E}^b,
\end{equation}
\begin{equation}
\vspace{-2mm}
\text{where}\quad{\cal E}^f=\sum_{n}{\left(x_n + \sum_{k=1}^{p}{a_k x_{n-k}}\right)^2}
\end{equation}
\begin{equation}
\vspace{-2mm}
\text{and}\quad{\cal E}^b=\sum_{n}{\left(x_n + \sum_{k=1}^{p}{a_k x_{n+k}}\right)^2}
\end{equation}
denote the forward and backward errors, respectively, $x_n$ denotes the current speech sample, and $a_k$ denotes the prediction coefficients.
The prediction coefficients can be computed by minimizing the combined error ($\partial{\cal E}/\partial{a_i}=0,\enskip 1\le i\le p$), which results in the following normal equations
\vspace{-2mm}
\begin{equation}
\sum_{k=1}^{p}c_{i,k}a_k=-c_{i,0}, \quad 1\le i\le p
\end{equation}
\begin{equation}
\text{where} \enskip c_{i,k}=\sum_{n}x_{n-i}x_{n-k} + \sum_{n}x_{n+i}x_{n+k}.
\end{equation}

Quasi-closed phase forward-backward (QCP-FB) analysis involves the use of FB analysis within the framework of WLP in order to combine the benefits of both techniques. WLP is computed using a temporal weighting function called the quasi-closed phase (QCP) function defined in \cite{Manu2014}. The QCP weighting function, which is computed automatically for every speech frame by first estimating glottal closure instants, has small values in the vicinity of glottal closure instants. Therefore, by using the QCP function in WLP analysis, the strong contribution of prediction error at glottal closure instants can be reduced, which yields improved formant estimates as reported in \cite{Manu2014}. The forward and backward errors are individually weighted using the QCP function. By denoting the QCP weighting function with $w_n$, the combined error to be minimized can be written as
\vspace{-2mm}
\begin{gather}
{\cal F}={\cal F}^f+{\cal F}^b,
\end{gather}
\vspace{-5mm}
\begin{gather}
\text{where}\quad{\cal F}^f=\sum_{n}{w_n\left(x_n + \sum_{k=1}^{p}{a_k x_{n-k}}\right)^2}
\end{gather}
\vspace{-4mm}
\begin{gather}
\text{and}\quad{\cal F}^b=\sum_{n}{w_n\left(x_n + \sum_{k=1}^{p}{a_k x_{n+k}}\right)^2}
\end{gather}
are the weighted forward and backward errors, respectively.
The resulting normal equations are given by
\vspace{-2mm}
\begin{gather}
\sum_{k=1}^{p}d_{i,k}a_k=-d_{i,0}, \quad 1\le i\le p \label{eq11}
\end{gather}
\vspace{-4mm}
\begin{gather}
\text{where} \enskip d_{i,k}=\sum_{n}w_nx_{n-i}x_{n-k} + \sum_{n}w_nx_{n+i}x_{n+k}. \label{eq12}
\end{gather}
An appropriate choice of range for the variable $n$ results in the autocorrelation or covariance methods for QCP-FB. In the current study, we use the covariance method in QCP-FB.

Both LP-COV and QCP-FB are computed using a frame length of 25 ms, a frame shift of 10 ms and an all-pole model order of $p$=13. Both methods are computed based on the covariance criterion using the rectangular window. Speech signals, sampled using 8 kHz, are pre-emphasized using an FIR filter ($P(z)=1-0.97z^{-1}$). The peaks in the spectrum are detected by convolving the spectrum with a Gaussian derivative window of width 100 Hz and picking the negative zero-crossings.

\begin{figure}
	\centering
	\includegraphics[width=\columnwidth,height=4.2cm, trim=1cm 0.5cm 0.5cm 0.5cm]{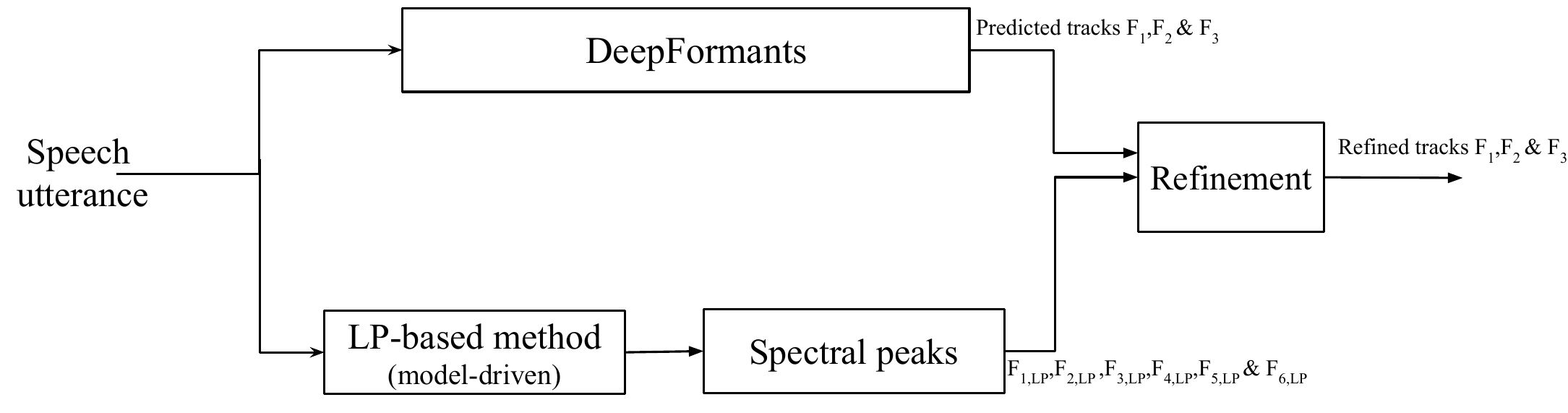}
	\caption{\label{fig:DF_LP_refinement_diagram}Illustration of refining the formant tracks computed by DeepFormants using an LP-based method. In the upper path, the three lowest formants are tracked frame-wise by DeepFormants. In the lower path, local spectral peaks, whose maximum number is six, are extracted from the all-pole spectrum computed frame-wise by an LP-based method. The outputs of both paths are used in the refinement as demonstrated in Figure 2 to define the refined formant tracks.}
\end{figure}

\subsection{Refining the formants tracked by DeepFormants using LP-based methods}
\label{sec:refining} 
\begin{figure}
	\centering
	\includegraphics[width=\columnwidth,height=6.5cm, trim=4cm 0.5cm 2cm 0.5cm]{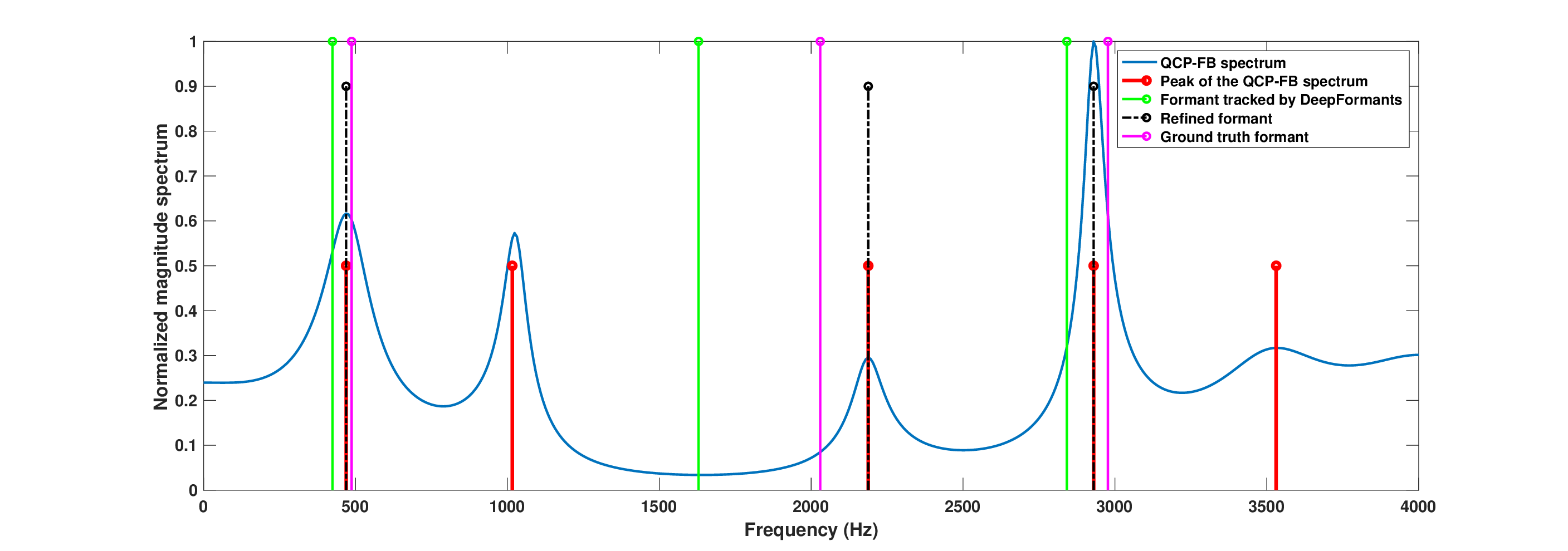}
	\caption{\label{fig:LP_Spectrum_DF_Refine}
 Illustration of refining the formants estimated by DeepFormants using the all-pole spectrum computed by an LP-based method (QCP-FB). In this example, the QCP-FB spectrum shows five local peaks (marked by red lines). The three formants predicted by DeepFormants are shown by green lines. The refined formants (marked by black lines) are those three local peaks in the QCP-FB spectrum that are closest to the formants predicted by DeepFormants.}
\end{figure}

\begin{figure}
\label{DFvsDFqcpfb}
	\centering
	\includegraphics[width=\columnwidth,height=14cm, trim=1cm 1cm 1cm 0cm]{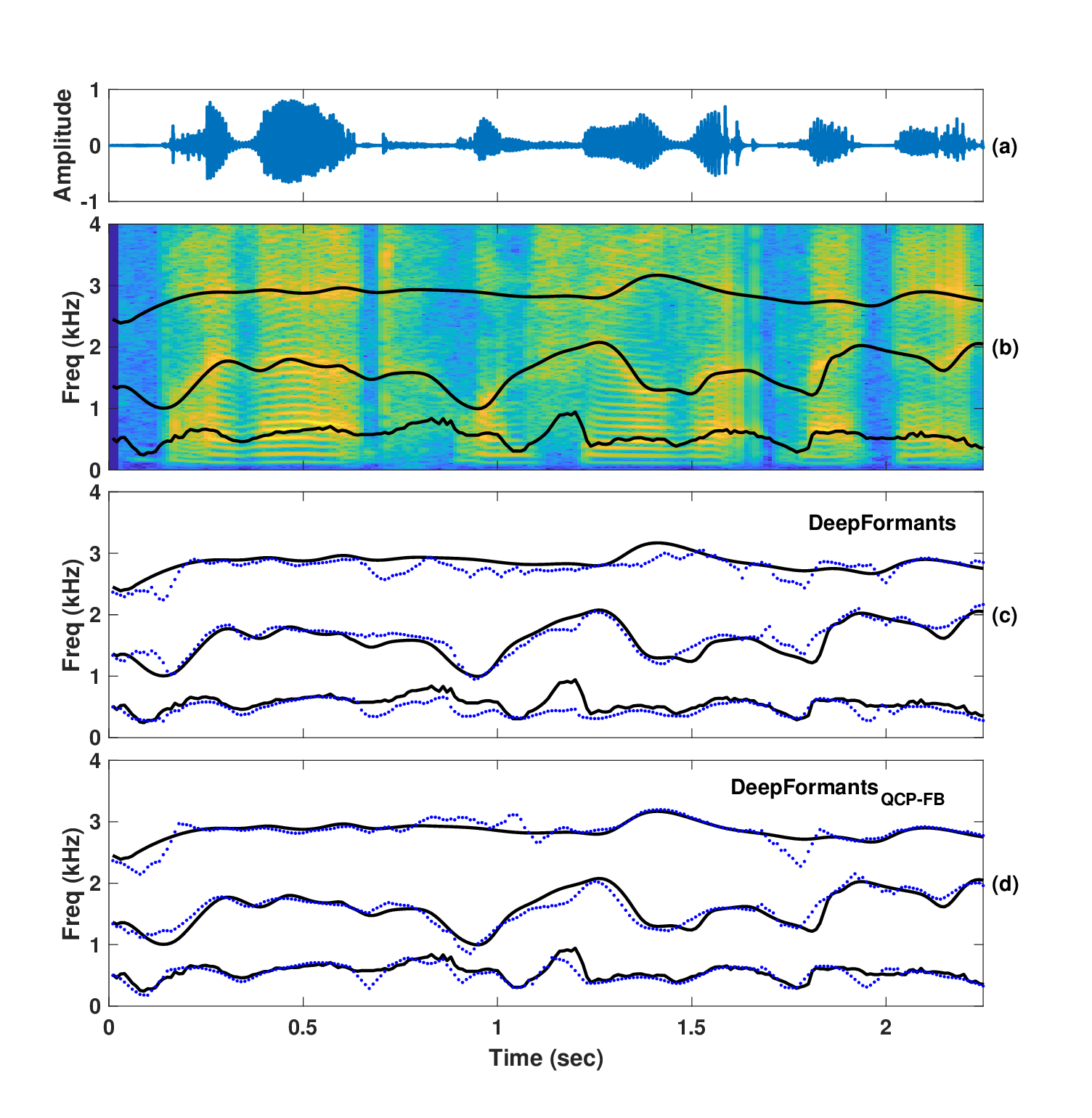}
\caption{\label{fig:DF_refine} {Formant frequencies of DeepFormants and DeepFormants${_{QCP-FB}}$ for a sentence produced by a male speaker: (a) the time-domain speech signal, (b) the narrowband spectrogram with reference ground truth formant contours, (c) the formant track estimates of DeepFormants, and (d) the formant track estimates of DeepFormants${_{QCP-FB}}$. The example was computed from sentence “By that, one feels that magnetic forces are as general as electrical forces.", from which a segment of 2.2 sec from the beginning is shown. Better performance of DeepFormants${_{QCP-FB}}$ compared to DeepFormants can be seen by comparing panels (c) and (d), for example, in tracking $F_1$ (between 0.6 and 1.2 sec) and in tracking $F_3$ (between 1.4 and 1.5 sec).}}
\end{figure}

The proposed refinement approach, which is shown as a flow diagram in Fig.~\ref{fig:DF_LP_refinement_diagram}, modifies the formants tracked by DeepFormants using a procedure consisting of the following parts. First, DeepFormants maps the acoustical features that are computed frame-wise as described in Section \ref{sec:deepformants} into preliminary contours of $F_1$, $F_2$, and $F_3$. Second, the LP-based all-pole spectral model (LP-COV or QCP-FB) is computed from each input frame and local peaks of the all-pole spectrum are determined. Note that with the all-pole model order $p$=13, both LP-COV and QCP-FB can maximally show six local peaks in their spectra. Third, each of the three preliminary formants predicted by DeepFormants are replaced in all frames with the local peak of the all-pole spectrum that is closest to the corresponding formant predicted DeepFormants. A graphical demonstration of the procedure to select the peaks of the all-pole spectrum is shown in Fig.~\ref{fig:LP_Spectrum_DF_Refine}. In the remaining sections, the refined DeepFormants tracker using LP-COV and QCP-FB in computing spectral peaks is denoted by
DeepFormants${_{LP-COV}}$ and DeepFormants${_{QCP-FB}}$, respectively.
An illustration of formant frequencies of DeepFormants and DeepFormants${_{QCP-FB}}$ for an utterance produced by a male speaker is shown in Fig.~\ref{fig:DF_refine}. {It can be seen from the figure that DeepFormants${_{QCP-FB}}$ is able to match the ground truth formant contours more closely than DeepFormants. }

\section{Experiments}
\label{sec:experiments} 
\subsection{Database}
\label{sec:database} 
Performance of the different formant trackers was evaluated using the VTR database, which is one of the most widely used speech databases in the areas of formant estimation and tracking~\cite{lideng2006}. The test set of the database was used for the evaluation. This data consists of 192 utterances produced by 8 female and 16 male speakers, each pronouncing eight utterances. The duration of each utterance varies between two and five seconds. The ground truth formant frequencies have been derived using a semi-supervised LP-based method \cite{lideng2004}. The first three formant frequencies ($F_1$, $F_2$, and $F_3$) have been corrected manually using spectrograms. The ground truth values for formants are provided for every 10 ms interval. 

\subsection{Performance metrics}
\label{sec:metrics} 
The formant tracking performance was evaluated using two metrics, the formant detection rate (FDR) and the formant estimation error (FEE), which have been used previously in formant estimation and tracking studies \cite{QCPFB_JASA, prasad21_interspeech}. FDR is measured in terms of the percentage of frames where a formant is hypothesized within a specified deviation from the ground truth.
The FDR for the $i^{th}$ formant over $K$ analysis frames is computed as
\begin{align}
D_{i} &=  \frac{1}{K}\sum_{n=1}^{K}{I(\Delta F_{i,n})} \\
I(\Delta F_{i,n}) &= \left\{\begin{array}{ll} 1 &\quad\text{if} \left({\Delta F_{i,n}}/{F_{i,n}} < \tau_r \quad \& \quad \Delta F_{i,n} < \tau_a\right) \\ 0 &\quad \text{otherwise}\end{array}\right. \label{eq:fdr}
\end{align}
where $I(.)$ denotes a binary formant detector function and $\Delta{F_{i,n}}=|F_{i,n}-\hat{F}_{i,n}|$ is the absolute deviation of the hypothesized formant frequency $\hat{F}_{i,n}$ for $i^{th}$ formant at the $n^{th}$ frame from the reference ground truth $F_{i,n}$.
The thresholds $\tau_r$ and $\tau_a$ denote the relative deviation and absolute deviation, respectively. As in \cite{QCPFB_JASA,gowda2020time}, these parameters were set as  $\tau_r$ = 30\% and $\tau_a$ = 300 Hz. FEE is measured in terms of the average absolute deviation of the hypothesized formants from the ground truth.
The FEE for the $i^{th}$ formant over $K$ analysis frames is computed as
\begin{equation}
E_i=\frac{1}{K}\sum_{n=1}^{K}{\Delta F_{i,n}}.
\label{eq:fee}
\end{equation}
The better the performance of a formant tracker, the larger the value of FDR and the smaller the value of FEE will be. The FDR and FEE values were computed in the current study for frames which were in the particular phonetic category of interest (see Section \ref{sec:results}).

\subsection{Reference formant tracking methods}
\label{sec:reference-methods} 
The reference trackers studied in the current article include the following six formant trackers: (1) the PRAAT algorithm based on the BURG method in LP analysis ~\cite{praat2001} (denoted as PBURG), (2) the adaptive filter bank (AFB) -based formant tracking algorithm proposed in~\cite{Mustafa2006} (denoted as MUST), (3) the Wavesurfer tracker ~\cite{wavesurfer2000} based on the autocorrelation method in LP (denoted as WSURF-0), (4) the Wavesurfer tracker ~\cite{wavesurfer2000} based on the covariance method in LP (denoted as WSURF-1), (5) the Kalman filtering-based tracker proposed in \cite{Mehta2012} (denoted as KARMA), and (6) the default DL-based DeepFormants tracker proposed in \cite{Dissen2019} (denoted as DeepFormants). The first five represent classic formant trackers and they were allowed to track three formants from the underlying spectrum at a frame rate of 100 Hz.

\begin{table}
\centering
\caption{\label{tab:ftrack1} Formant tracking results obtained using the evaluation dataset (vowels, diphthongs, semivowels) of the VTR corpus. Formants were tracked using five conventional formant trackers (PBURG, MUST, WSURF-0, WSURF-1, and KARMA), DeepFormants and two refined versions of DeepFormants (DeepFormants${_{LP-COV}}$ and DeepFormants${_{QCP-FB}}$). FDR denotes formant detection rate and FEE denotes formant estimation error.}
\vspace{0.2cm}
\resizebox{16cm}{4.5cm}{
\begin{tabular}{|c||c|c|c||c|c|c|} \hline
& \multicolumn{3}{c||}{FDR (\%)} & \multicolumn{3}{c|}{FEE (Hz)} \\\cline{2-7}
Method ~&~ $F_1$ ~&~ $F_2$ ~&~ $F_3$ ~&~ $\delta F_1$ ~&~ $\delta F_2$ ~&~ $\delta F_3$ ~\\\hline\hline
PBURG & 86.0 & 70.0 & 63.1 & 88 & 268 & 340 \\\hline
MUST & 81.1 & { 86.3} & 76.9 & 91 & { 152} & 230 \\\hline
WSURF-0 & 84.1 & 78.2 & 77.3 & 93 & 239 & 245 \\\hline
WSURF-1 & 86.6 & 82.7 & 80.8 & 87 & 223 & 228 \\\hline
KARMA &91.5 &89.4 &74.7 & { 62} & 146 & 250 \\\hline
DeepFormants  & 91.7 & 92.3 & 89.7 & 85 & 120 & 143  \\\hline 

DeepFormants${_{LP-COV}}$   & 93.0 & 93.8 & 90.6 & 62 & 109 & 138 \\ \hline 

DeepFormants${_{QCP-FB}}$ & 93.9 & 94.4 & 92.6 & 60 & 103 & 119\\ \hline
\end{tabular}}
\end{table}

\section{Results}
\label{sec:results} 
{ As the first experiment, we compared all trackers described in Section \ref{sec:reference-methods} by combining the vowels, diphthongs, and semivowels of the VTR corpus into a joint evaluation dataset. The obtained results are shown for the five traditional trackers, DeepFormants, and for the two refined DeepFormants trackers in Table~\ref{tab:ftrack1}. 
The data shows that the formant tracking performance obtained by the DL-based trackers (DeepFormants,  DeepFormants${_{LP-COV}}$, and DeepFormants${_{QCP-FB}}$) were clearly better compared to the traditional trackers. From the traditional trackers, KARMA was able to track $F_1$ with performance that was comparable to that of the three DL-based trackers, but KARMA's performance for the other two formants, particularly  $F_3$, was lower. By comparing the three DL-based trackers, a consistent trend can be seen in both metrics: the two refined versions of DeepFormant (DeepFormants${_{LP-COV}}$ and DeepFormants${_{QCP-FB}}$) were better than DeepFormant, and the refinement based on QCP-FB outperformed the refinement based on LP-COV. It can also be observed that DeepFormant gave an estimation error for $F_1$ that was more than 20 Hz larger than that of the two refined trackers
, but also more than 20 Hz larger than that of KARMA. }

\begin{table}
\centering
\vspace{-2.5cm}
\caption{\label{tab:ftrack2} { Formant tracking results obtained by dividing the evaluation dataset of the VTR corpus into different phonetic categories. Formants were tracked using KARMA, DeepFormants, DeepFormants${_{LP-COV}}$ and DeepFormants${_{QCP-FB}}$. FDR denotes formant detection rate and FEE denotes formant estimation error.}}
\vspace{0.1cm}
\resizebox{15cm}{11.5cm}{
\begin{tabular}{|c||c|c|c||c|c|c|}
\hline
 & \multicolumn{3}{c||}{FDR (\%)} & \multicolumn{3}{c|}{FEE (Hz)} \\\cline{2-7}
Method ~&~ $F_1$ ~&~ $F_2$ ~&~ $F_3$ ~&~ $\delta F_1$ ~&~ $\delta F_2$ ~&~ $\delta F_3$ ~\\\hline\hline
\multicolumn{7}{|c|}{\bf Vowels}\\\hline
KARMA  &92.6 & 89.0 & 74.5 & 57 & 150 & 251   \\ \hline
DeepFormants  &92.7 & 93.7 & 91.0 & 82 & 113 & 135 \\\hline
DeepFormants${_{LP-COV}}$  &94.2 & 95.2 & 91.1 & 56 & 97 & 134  \\ \hline
DeepFormants${_{QCP-FB}}$  &94.8 & 95.7 & 93.7 & 56 & 94  & 109 \\\hline \hline

\multicolumn{7}{|c|}{\bf Diphthongs}\\\hline
KARMA  & 92.5 &92.3  &76.5  & 63 &129   &240  \\ \hline
DeepFormants  &93.2 & 93.8 & 90.6 & 85 & 112 & 133 \\\hline 
DeepFormants${_{LP-COV}}$  &93.0 & 93.8 & 90.6 & 62 & 109 & 138 \\ \hline
DeepFormants${_{QCP-FB}}$     &94.7 & 95.6 & 94.0 & 62 & 99 & 108 \\\hline \hline

\multicolumn{7}{|c|}{\bf Semivowels}\\\hline
KARMA  &86.9  &86.9  &73.6  &76  &155   &258  \\ \hline
DeepFormants &87.0 & 86.1 & 84.4 & 96 & 148 & 176 \\\hline 
DeepFormants${_{LP-COV}}$  &88.9 & 88.4 & 85.4 & 73 & 147 & 179  \\ \hline
DeepFormants${_{QCP-FB}}$      &90.1 & 89.1 & 87.6 & 72 & 136 & 160 \\\hline \hline

\multicolumn{7}{|c|}{\bf Nasals}\\\hline
KARMA  &82.1 & 80.7 & 75.0 & 90 & 214 & 241  \\ \hline
DeepFormants    &79.9 & 80.3 & 86.4 & 97 & 181 & 160 \\\hline 
DeepFormants${_{LP-COV}}$  &85.9 & 84.9 & 86.9 & 76 & 167 & 161  \\ \hline
DeepFormants${_{QCP-FB}}$     &86.2 & 85.2 & 87.9 & 76 & 159 & 149 \\\hline \hline 

\multicolumn{7}{|c|}{\bf Fricatives}\\\hline
KARMA  &56.4 & 85.9 & 73.8 & 191 & 160 & 244  \\ \hline
DeepFormants  &67.9 & 87.0 & 83.4 & 138 & 159 & 174 \\\hline
DeepFormants${_{LP-COV}}$  &61.5 & 86.0 & 83.2 & 161 & 156 & 171  \\ \hline
DeepFormants${_{QCP-FB}}$  &61.5 & 87.3 & 84.0 & 164 & 149 & 165 \\\hline \hline

\multicolumn{7}{|c|}{\bf Voice Bars}\\\hline
KARMA  &71.5 & 87.8 & 75.0 & 92 & 151 & 256   \\ \hline
DeepFormants  &79.1 & 86.4 & 84.8 & 81 & 156 & 163 \\\hline
DeepFormants${_{LP-COV}}$  &73.6 & 89.4 & 81.4 & 83 & 138 & 185  \\ \hline
DeepFormants${_{QCP-FB}}$      &73.9 & 90.9 & 83.8 & 83 & 132 & 166 \\\hline \hline 

\multicolumn{7}{|c|}{\bf Stops}\\\hline
KARMA  & 65.2 & 86.4 & 72.4 & 157 & 159 & 255  \\ \hline
DeepFormants      &63.2 & 88.5 & 82.6 & 152 & 151 & 176\\\hline
DeepFormants${_{LP-COV}}$  &69.5 & 87.9 & 82.7 & 139 & 148 & 182  \\ \hline
DeepFormants${_{QCP-FB}}$      &69.0 & 87.3 & 83.7 & 140 & 150 & 173\\\hline 
\end{tabular}}
\end{table}

{As the second experiment, formant tracking performance of the four best trackers reported in  Table~\ref{tab:ftrack1} (KARMA, DeepFormants, DeepFormants${_{LP-COV}}$, and DeepFormants${_{QCP-FB}}$) was further investigated by using evaluation data that consisted of several fine-grained phonetic categories. The corresponding results are shown in Table~\ref{tab:ftrack2}. The data in Table~\ref{tab:ftrack2} is in line with previous studies \cite{Dissen2019, Dai2020IS} indicating that formant tracking performance was highest for vowels, semivowels, and diphthongs, but performance reduced for other categories, particularly for fricatives and stops. By comparing the four trackers for vowels, diphthongs, semivowels, and nasals, it can be seen that the two refined trackers gave smaller estimation errors than KARMA and DeepFormants for all formants except in a few individual cases (e.g., in diphthongs, a slightly smaller FEE value in $F_3$ was shown by DeepFormants compared to DeepFormants${_{LP-COV}}$). For fricatives, voice bars, and stops, the two refined trackers gave a smaller FEE value than KARMA consistently for all three formants. DeepFormants, however, was the best tracker in a few combinations of phonetic category and formant. In particular, the FEE value given by DeepFormants in tracking $F_1$ of fricatives was smaller than the corresponding error given by the other three trackers.

\begin{table}
\centering
\vspace{-2.5cm}
\caption{\label{tab:noisecompare} 
{Formant tracking results obtained by degrading the evaluation dataset (vowels, diphthongs, semivowels) of the VTR corpus with babble and white noise at SNR levels of 20 dB, 10 dB, and 5 dB. Formants were tracked using KARMA, DeepFormants, DeepFormants${_{LP-COV}}$, and DeepFormants${_{QCP-FB}}$. FDR denotes formant detection rate and FEE denotes formant estimation error.}}
\vspace{0.2cm}
\resizebox{15cm}{11.5cm}{
\begin{tabular}{|c||c|c|c||c|c|c|}
\hline
 & \multicolumn{3}{|c||}{FDR (\%)} & \multicolumn{3}{|c|}{FEE (Hz)} \\\cline{2-7}
Method & $F_1$ & $F_2$ & $F_3$ & $\delta F_1$ & $\delta F_2$ & $\delta F_3$\\\hline

\multicolumn{7}{c}{}\\\hline

\multicolumn{7}{|c|}{{\bf Babble at 20 dB}} \\\hline
KARMA        &{91.7} & 88.0 & 74.2 & {61} & 153 & 248  \\\hline
DeepFormants        &91.3 & 91.7 & 87.1 & 90 & 118 & 155   \\\hline
DeepFormants${_{LP-COV}}$      &92.5 & 92.1 & 89.7 & 62 & 117 & 139\\\hline
DeepFormants${_{QCP-FB}}$      &93.5 & 92.3 & 91.2 & 61 & 114 & 125  \\\hline

\multicolumn{7}{|c|}{{\bf Babble at 10 dB}} \\\hline
KARMA        &90.3 & 83.8 & 71.8 & {65} & 176 & 246  \\\hline
DeepFormants        &{91.1} & 86.6 & 81.7 & 88 & 146 & 183   \\\hline
DeepFormants${_{LP-COV}}$      &90.7 & 86.4 & 83.9 & 65 & 150 & 170 \\\hline
DeepFormants${_{QCP-FB}}$      &91.3 & 86.6 & 85.5 & 63 & 147 & 158  \\\hline

\multicolumn{7}{|c|}{{\bf Babble at 5 dB}} \\\hline
KARMA        &88.2 & 78.9 & 68.7 & {71} & 201 & 260   \\\hline
DeepFormants        &{89.8} & 81.4 & 76.1 & 90 & 177 & {209}   \\\hline
DeepFormants${_{LP-COV}}$      &88.6 & 80.5 & 78.0 & 68 & 185 & 201\\\hline
DeepFormants${_{QCP-FB}}$      &89.0 & 81.0 & 79.0 & 67 & 180 & 192  \\\hline

\multicolumn{7}{c}{}\\\hline

\multicolumn{7}{|c|}{{\bf White at 20 dB}} \\\hline
KARMA        &90.4 & 87.6 & 73.6 & {64} & 151 & 241 \\\hline
DeepFormants        &90.1 & 90.4 & {84.4} & 95 & {126} & {168}   \\\hline
DeepFormants${_{LP-COV}}$      &92.6 & 91.1 & 85.0 & 61 & 126 & 170 \\\hline 
DeepFormants${_{QCP-FB}}$      &93.5 & 91.8 & 86.9 & 60 & 120 & 153   \\\hline

\multicolumn{7}{|c|}{{\bf White at 10 dB}} \\\hline
KARMA        &86.2 & 80.1 & 68.8 & 76 & 191 & 257 \\\hline
DeepFormants        &89.8 & 80.8 & {71.6} & 99 & 184 & {239}   \\\hline
DeepFormants${_{LP-COV}}$      &91.1 & 83.0 & 73.0 & 63 & 176 & 242 \\\hline
DeepFormants${_{QCP-FB}}$      &92.1 & 83.4 & 74.2 & 62 & 170 & 228 \\\hline

\multicolumn{7}{|c|}{{\bf White at 5 dB}} \\\hline
KARMA        &80.1 & 72.5 & 64.0 & {92} & 233 & 279\\\hline
DeepFormants        &{89.2} & 71.7 & {64.5} & 101 & 239 & {274}  \\\hline
DeepFormants${_{LP-COV}}$      &88.0 & 74.4 & 66.2 & 71 & 230 & 280\\\hline
DeepFormants${_{QCP-FB}}$      & 88.8 & 74.7 & 66.5 & 69 & 226 & 271 \\\hline
\end{tabular}
}
\end{table}

As the third experiment, noise robustness of the four best trackers reported in  Table~\ref{tab:ftrack1} (KARMA, DeepFormants, DeepFormants${_{LP-COV}}$, and DeepFormants${_{QCP-FB}}$) was studied by corrupting the (clean) speech input of the VTR corpus with additive noise. The noise corruption was done using two types of noise (babble and white) and three signal-to-noise ratios (SNRs) (20 dB, 10 dB, and 5 dB). The experiment was conducted in the similar manner as the first experiment (reported in Table~\ref{tab:ftrack1}) by using as evaluation data the vowels, diphthongs, and semivowels of the VTR corpus. The results of the corresponding tracking experiments are reported in Table~\ref{tab:noisecompare}. As the main results, the following observations can be made about the FEE values shown in the table. First, KARMA gave clearly larger estimation errors than the other three trackers for $F_2$ and $F_3$ in all noise-corruption categories (except for white noise with SNR=5 dB, for which the FEE metrics were almost the same for all trackers both in $F_2$ and $F_3$). Second, DeepFormants gave the largest estimation error in tracking $F_1$ in all noise-corruption categories. Interestingly, however,  the performance of DeepFormants in $F_1$ tracking did not drop as as much as that of, for example, KARMA when the amount of noise was increased by changing SNR from 20 dB to 5 dB. The best robustness against noise was shown by DeepFormants${_{QCP-FB}}$ as can be seen, for example, by comparing the FEE value of $F_1$ between the clean condition (i.e., Table~\ref{tab:ftrack1}) and the most severe noise condition (i.e., white noise at 5 dB in Table~\ref{tab:noisecompare}). By comparing these two tables, it can be seen that for  DeepFormants${_{QCP-FB}}$ the value of FEE rose from 60 Hz to 69 Hz, whereas the corresponding rise for DeepFormants was from 85 Hz to 101 Hz, and for KARMA from 62 Hz to 92 Hz.}
\clearpage
\section{{Discussion and conclusions}}
\vspace{-0.3cm}
\label{sec:conclusions} 
{Formant tracking was studied in this article based on refining the formant contours predicted
by an existing modern data-driven formant tracker, DeepFormants. In the studied approach, the trained NN model of the DeepFormants tracker first maps input speech frames to the preliminary contours of $F_1$, $F_2$, and $F_3$. The predicted formants are then replaced frame-wise by the local peaks in the all-pole spectra computed in a model-driven manner by an LP-based method from the same speech frames. As an LP method, one conventional method (LP-COV) and one recently developed algorithm (QCP-FB) were used. The proposed refinement technique was compared in formant tracking to the original version of DeepFormants and to five classic trackers. In the first experiment, altogether eight trackers (five classical trackers, DeepFormants, DeepFormants${_{LP-COV}}$, and DeepFormants${_{QCP-FB}}$) were compared using evaluation data that consisted of the vowels, diphthongs, and semivowels of the VTR corpus. The results showed that the three DeepFormants-based trackers outperformed all five classic trackers, and that DeepFormants${_{QCP-FB}}$ was the best tracker. In the second experiment, four best trackers of the first experiment were compared 
by dividing the evaluation data into seven phonetic categories. The results indicated that the refined DeepFormants${_{QCP-FB}}$ tracker showed a consistent performance improvement over the original DeepFormants tracker in all three formants for vowels, diphthongs, semivowels, and nasals. Particularly for fricatives, however, tracking of $F_1$ showed lower performance for DeepFormants${_{QCP-FB}}$ compared to DeepFormants. Finally, in the third experiment, robustness of formant tracking with respect to noise was evaluated by comparing the same trackers that were included in the second experiment. The results of this experiment indicated that DeepFormants${_{QCP-FB}}$ showed the best resilience to noise.}

{ The results summarized above suggest that the proposed idea to refine the formants tracked by DeepFormants using the formants predicted by a model-driven, LP-based signal processing approach results in an improved accuracy and noise robustness in formant tracking. The observed performance degradation of DeepFormants${_{QCP-FB}}$ in tracking formants of fricatives is due to the aperiodic nature of the excitation of the speech production mechanism in these sounds. For speech sounds with aperiodic glottal excitations, there are namely no clear glottal closure instants, and therefore QCP-FB analysis is not able to improve the estimation of the vocal tract by reducing the effect of glottal closure instants. The improved noise robustness of DeepFormants${_{QCP-FB}}$ can in turn be explained by two issues. First, QCP-FB analysis estimates vocal tract resonances by emphasizing the contribution of speech waveform samples that occur after glottal closure \cite{QCPFB_JASA}. These speech samples are more robust against noise because their amplitude level is larger than in other parts of the glottal cycle. Therefore, QCP-FB analysis inherently utilizes speech samples that are more resilient to additive noise, which leads to improved formant tracking by DeepFormants${_{QCP-FB}}$. Second, the key concept of DeepFormants${_{QCP-FB}}$, the combination of model-driven and data-driven approaches, in formant tracking helps in improving noise robustness of formant tracking. In purely data-driven trackers, such as DeepFormants, there is always the risk of having a mismatch between the system training and test stages. The DeepFormants tracker has been namely trained with clean speech, and therefore it is understandable that its performance might drop when the tracker is tested in conditions that have not been seen by the tracker's NN models in the system training stage. In DeepFormants${_{QCP-FB}}$, however, the other part of the combination, the model-driven QCP-FB technique, is free of any data learning stage, and therefore the tracker will be more resilient to mismatch between the training and test conditions. 
}

{ In reporting results of formant tracking experiments (e.g., \cite{Dissen2019, Lilley2021, Shrem2022}), authors typically compare their proposed method with known reference techniques by using  absolute FEE measure as metrics (in Hz, defined Eq.~\ref{eq:fee}), which was also used in this study. In some studies (e.g., \cite{Vijayan2019}), results are reported using the mean absolute deviation (MAD), which is the relative error (in \%) between the tracked formant and its reference value. (Note that MAD is called the mean absolute percentage error in some studies \cite{Dai2020, Dai2020IS}). What is, however, left undiscussed by authors of the study area is the question whether the tracking metrics obtained can be considered good enough when the underlying formant tracker is in use, for example, in phonetic studies of natural speech.  This question about the goodness of metrics is obviously difficult to answer because formant tracking can be used in tasks that are of different requirements in terms of how large errors formant tracking is allowed to show. As an example, in identification of vowels based on their tracked $F_1$ and $F_2$ values, a larger error may be tolerated for vowels that are at the corners of the vowel triangle (i.e., /u/, /i/, and /a/) compared to those that are in the middle (e.g., \textipa{/\oe/} and \textipa{/\textepsilon/}). In order to shed light on this problematic question about the goodness of metrics, we propose that the assessment of the metrics shown by a formant tracker could be done by comparing the tracker's metrics with the results reported in human perception studies on formants. There are namely many investigations that have been done since the 1950's to study human perceptual thresholds in formant distinction (e.g., \cite{Flanagan55, Gagne88}). The results of these studies were summarized by Kewley-Port and Watson in \cite{Kewley-Port}, and their conclusion was that humans are able to distinguish changes in $F_1$ when its value is altered by 3\%-10\% (corresponding to MAD values between 3\% and 10\%). Achieving MAD values of this small magnitude even for one phonetic category (e.g., vowels which were considered in \cite{Kewley-Port}) is obviously a tough requirement in automatic formant tracking. In order to demonstrate this for the trackers of the current study, we computed MAD values for DeepFormants and DeepFormants${_{QCP-FB}}$ in tracking $F_1$ from vowels and diphthongs. For vowels, MAD values of 15.7\% and 11.4\% were obtained by DeepFormants and DeepFormants${_{QCP-FB}}$, respectively. For diphthongs, the corresponding values were 15.3\% and 11.6\%. Hence, the performance of even the best trackers is still many percentage units lower compared to human performance in detection of noticeable changes in $F_1$.  It is undoubtedly ambitious to set the goal in formant tracking 
to metrics values that correspond to the thresholds reported in formant distinction by humans, and this requirement seems not be possible to be achieved yet even by the latest data-driven trackers. However, we hope that the introduction of this goal motivates the development of new formant trackers, and we argue the goal is in line with recent progress in other areas of speech technology, such as text-to-speech synthesis \cite{Tacotron2} and automatic speech recognition \cite{Microsoft}, where performance of machine is approaching human performance.
} 

{In conclusion,} the study showed that the two groups of formant estimation techniques that have been used in formant trackers, the conventional model-based approach based on parametric all-pole modeling and the modern data-driven approach, should not be seen as alternatives to each other. Instead they can be used together to enhance the accuracy of data-driven trackers. Accuracy improvement can be implemented easily by plugging an LP-based formant estimation method into an existing data-driven tracker without re-training the tracker's NNs. { Future studies will be conducted to analyze the system performance when the test data is taken from other annotated databases than the VTR corpus, which was used in the training of DeepFormants}.

\section*{Acknowledgments}
This work has been funded by the Academy of Finland (project no 330139) and Aalto University (the Ministry of Education and Culture’s Global Program Pilots for India).


\end{document}